\documentclass[a4paper,11pt]{article}
\pdfoutput=1 

\usepackage{jheppub}

\usepackage[mmddyyyy,hhmmss]{datetime}

\usepackage{amsmath,graphicx,amssymb,subfigure,bbm,comment}
\usepackage[all]{xy}

\allowdisplaybreaks

\title{Finite size effects in classical string solutions of the Schr\"odinger geometry}

\author{Dimitrios Zoakos}

\affiliation{Department of Physics, National and Kapodistrian University of Athens, 15784 Athens, Greece.}
\affiliation{Hellenic American University, 436 Amherst st, Nashua, NH 03063 USA.}

\emailAdd{zoakos@gmail.com}

\abstract{We study finite size corrections to the semiclassical string solutions of the Schr\"odinger spacetime.
We compute the leading order exponential corrections to the infinite size dispersion relation of the single spin giant magnon 
and of the single spin single spike solutions. The solutions live in a $S^3$ subspace of the five-sphere and extent in the 
Schr\"odinger part of the metric. In the limit of zero deformation the finite size dispersion relations flow 
to the undeformed $AdS_5 \times S^5$  counterparts and in the infinite size limit the correction term vanishes and the known infinite size dispersion relations are obtained.}

\begin{document}
\maketitle
\flushbottom

\section{Introduction}

The powerful tool of integrability (for a review see \cite{Beisert:2010jr}) is behind all the recent advances in understanding the
celebrated gauge-gravity correspondence \cite{Maldacena:1997re} 
(for a set of pedagogical introductions see \cite{Ramallo:2013bua,Edelstein:2009iv}) in the planar limit. 
On the gauge theory side, using Bethe ansatz techniques \cite{Minahan:2002ve},  the matrix of the 
anomalous dimensions can be diagonalized through the mapping to the Hamiltonian of an integrable spin chain.
The magnons are the fundamental single particle excitations that propagate on the BMN vacuum and integrability 
implies that every scattering process factorizes in a scattering between two magnons. 
The role of integrability is pivotal not only in determining the spectrum of ${\cal N}=4$ SYM at any value of the coupling 
but also in the computation of higher point correlation functions (see e.g. 
\cite{Zarembo:2010rr, Costa:2010rz,Georgiou:2010an, Georgiou:2011qk, Escobedo:2010xs,Kazama:2016cfl})

The strong coupling dual of the magnons are semi-classical string solutions on $AdS_5\times S^5$, the so-called giant magnons 
\cite{Hofman:2006xt,Minahan:2006bd,Chen:2006gea,Benvenuti:2008bd}.
They are open strings moving in a subspace of the five-sphere with finite angular extent. 
Another class of classical solutions that we will focus our attention in the current paper is the single spikes 
\cite{Kruczenski:2006pk,Ishizeki:2007we}. They have finite angular amplitude and wind infinitely many times around 
an angular direction. 

Over the years there is a lot of attention on integrable deformations of the original AdS/CFT construction. 
Recent activity is coming from the null dipole deformation of the ${\cal N}=4$ SYM and more precisely 
on the integrability issues that arise, as they were discussed \cite{Guica:2017mtd}.
There, a test of the Schr\"odinger holography is provided by matching the
anomalous dimensions of long gauge theory operators with the prediction at strong coupling of certain BMN-like strings 
(see also \cite{Ouyang:2017yko}). 

In \cite{Georgiou:2017pvi} semi-classical string solutions living in the Schr\"odinger spacetime were constructed. 
They are the counterparts of the giant magnon and the single spike solutions of the undeformed $AdS_5 \times S^5$, 
since in the limit of zero deformation they flow to the ordinary giant magnon and single spike of the 
undeformed background. The solutions live in an $S^3$ subspace of the five-sphere in which the $B$-field has 
non-zero components. Furthermore, the solutions are not point like but extend in the  Schr\"odinger  part of the 
metric.\footnote{Giant-magnon like solutions with a different dispersion relation 
were studied in \cite{Ahn:2017bio}, while giant magnons and spiky strings living on the 
Schr\"odinger $Sch_5 \times T^{1,1}$ and the corresponding dispersion relations were studied in  
\cite{Golubtsova:2020fpm}.}

In \cite{Georgiou:2018zkt} three-point correlation functions involving two {\it heavy} operators and a {\it light} one 
were calculated using holography in a Schr\"odinger background. 
These are the first results in the literature of three-point function computations 
involving extended string solutions.
In \cite{Georgiou:2019lqh} the pp-wave geometry of the Schr\"{o}dinger background was constructed. 
The spectrum of the bosonic excitations was derived and compelling agreement with the giant magnon dispersion relation, 
previously obtained in \cite{Georgiou:2017pvi}, was found.
 In  \cite{Georgiou:2020qnh} the giant graviton solution of the Schr\"{o}dinger pp-wave geometry \cite{Georgiou:2019lqh} 
was constructed. The solution exhibits an intriguing behavior as the deformation 
parameter varies,  which is suggestive of  a spontaneous breaking of conformal invariance. 
Finally, pulsating strings solutions in the Schr\"{o}dinger background were constructed in \cite{Dimov:2019koi}.

The aim of this work is to continue in the direction of further studying the gauge/gravity correspondence, 
in order to relate the non-supersymmetric Schr\"{o}dinger background to its dual null dipole-deformed ${\cal N}=4$ SYM 
\cite{Maldacena:2008wh,Herzog:2008wg,Adams:2008wt}. To that extent, in this paper we will further elaborate on the
classical string solutions that were constructed in \cite{Georgiou:2017pvi}.
We will calculate the finite size (exponential) corrections to the infinite size dispersion relation of the single spin giant magnon 
and of the single spin single spike solutions.
The solutions live in a $S^3$ subspace of the five-sphere, extent in the 
Schr\"odinger part of the metric and in the limit of zero deformation the finite size dispersion relations flow 
to the undeformed $AdS_5 \times S^5$  counterparts \cite{Arutyunov:2006gs, Ahn:2008sk}.

The paper is organized as follows: In section \ref{ClassicalSolutions} we review the single spin giant magnon and the 
single spin single spike solutions in the Schrodinger background. The ansatz for the solutions was introduced in 
\cite{Georgiou:2018zkt,Georgiou:2017pvi} but the boundary conditions are revised in order to accommodate the 
finite size corrections. In section \ref{section-GM} we focus on the giant magnon case and we calculate the explicit 
expressions for the conserved charges. Afterwards, we expand the dispersion relation around the infinite size limit and the 
zero deformation limit. 
In section \ref{section-SS} we focus on the single spike case and after calculating the explicit 
expressions for the conserved charges we expand the dispersion relation around the infinite size limit.
We conclude the paper in section \ref{conclusions}. In appendixes \ref{appendix-1} and \ref{appendix-2} we present 
details of the computations of the main text.


\section{Classical string solutions on $Sch_5\times S^3$}
\label{ClassicalSolutions}

In this section we will review the single spin giant magnon and the single spin single spike solutions 
in the Schrodinger  $Sch_5\times S^5$ background, that were initially presented in \cite{Georgiou:2018zkt,Georgiou:2017pvi}. 
More specifically, in \cite{Georgiou:2018zkt} the dispersion relations for the infinite size single spin giant magnon and 
the infinite size single spin single spike were presented, and it is those relations that in the next section we will generalize 
to include the finite size corrections. 

We consider the following consistent truncation of the 10d Schrodinger  $Sch_5\times S^5$ 
background on a sphere\footnote{With respect to the notation of \cite{Georgiou:2017pvi}
for the $S^3$, we have performed a change of variables: $\theta=2\,\eta$, $\psi=\varphi_1+\varphi_2$ and 
$\phi=\varphi_1-\varphi_2$. 
Details about the consistency of the truncation can be found in \cite{Georgiou:2017pvi}.} 
\begin{equation}
ds^2 = - \left(1+ \frac{\mu^2}{Z^4} \right) dT^2  +  \frac{1}{Z^2} \, \left(2 dT dV + dZ^2 \right) + 
d\eta^2 \, + \, \sin^2\eta \, d\varphi_1^2 + \cos^2\eta \, d\varphi_2^2 
\end{equation}
that is supplemented with a $B$-field
\begin{equation}
B \, = \,  \frac{\mu}{Z^2} \, d T \wedge \Big( \sin^2\eta \, d\varphi_1 + \cos^2\eta \, d\varphi_2 \Big) 
\end{equation}
where $\eta \in [0,\pi/2]$, $\varphi_1\in [0,2\pi)$ and $\varphi_2\in [0,2\pi)$ are the ranges of the variables along $S^3$.


\subsection{Ansatz and equations of motion}

We consider the following ansatz for both solutions
\begin{eqnarray} \label{ansatz}
&&
T\,=\, \kappa \,\tau\, ,
\hspace{1cm}
V \,=\,\alpha \, \tau \,  + \, V_y(y)\, ,  
\hspace{1cm}
Z \,=\, Z_0\, , 
\hspace{1cm}
\eta \,= \,\frac{1}{2} \,\theta_y(y)\, , 
\nonumber \\[7pt]
&& \hspace{1.5cm}
\varphi_1 \, = \, \omega \, \tau \,+ \frac{\kappa \, \mu}{Z_0^2}\, \sigma \, + \, \Psi_y(y)
\hspace{0.5cm} \& \hspace{0.5cm}
\varphi_2 \,=\, - \, \frac{\kappa \,  \mu}{Z_0^2} \,  \sigma
\end{eqnarray} 
where we have defined the variable $y$ as  
\begin{equation} \label{definition-y}
y\, \equiv \, c\,\sigma-d\,\tau \, . 
\end{equation}
The explicit expressions for the functions $V_y(y)$, $\theta_y(y)$ and $\Psi_y(y)$ that appear in the ansatz will be 
determined through the equations of motion and the Virasoro constraints, while $\kappa$, $\alpha$, $Z_0$,   
$\omega$ and $\mu$ are constants. 

Notice that even if we are considering single spin solutions, the string needs to move inside an $S^3$, rather than an $S^2$,
subspace of the 5-sphere. In order to have a dispersion relation that depends on the deformation, we are forced to 
{\it switch on} a third angle inside the 5-sphere. This is not something unusual when we study the motion of classical 
strings inside deformations of the $AdS$ space (e.g. see the case of the giant magnon inside the 
beta-deformed $AdS_5 \times S^5$ background \cite{Chu:2006ae}).

The equations for $V_y'(y)$ and $\Psi_y'(y)$ in terms of the the 
new function $u(y)$\footnote{The functions 
$\theta_y(y)$ and $u(y)$ are related as follows: $u \equiv \cos^2 \left(\frac{\theta_y}{2}\right)$.}
are given in appendix \ref{appendix-1},
while the equation of motion for $Z$ and one of the Virasoro constraints give us the following 
two conditions
\begin{equation} \label{constraintZ0}
A_T\, d \, Z_0^2 \, - \, 2 \, A_{\phi} \,c \, \mu \,+ \, \alpha \, c^2 \, - \, d^2 \, \kappa \, Z_0^2 \, = \, 0
\end{equation}
and
\begin{equation}
\label{virasoro}
\frac{A_{\phi} \, \omega}{d} \, - \, \frac{A_T \, \kappa}{2 \, d} \, +
\, \frac{1}{Z_0^2}\left(\frac{\alpha \,\kappa }{2} \, + \, \frac{A_{\phi} \, \kappa  \, \mu }{c}\right) \, = \, 0 \, .
\end{equation}
The equation of motion for $u(y) $ is coming from the other Virasoro constraint and it is
\begin{equation} \label{equation-u} 
\frac{(u')^2}{2}\,+\,{\cal W}(u) \,=\,0
\hspace{.7cm} \textrm{with} \hspace{.7cm}
{\cal W}(u)\,=\,-\,2 \, u \, \left(\beta_6\, u^2 + \beta_4\, u \,+ \, \beta_2   \right)
\end{equation}
where the constants $\beta_2$, $\beta_4$ \& $\beta_6$ are listed in appendix \ref{appendix-1}.
It is possible to rewrite equation \eqref{equation-u} in the following way
\begin{equation} \label{equation-u-v2}
u' \, = \, 2 \, \sqrt{\Big.|\beta_6| \, u \, \left( u_{p} \, - \, u \right) \, \left(u \, - \, u_{m} \right)}
\quad {\rm with} \quad
u_{m} < u < u_{p} 
\end{equation}
where
\begin{equation} \label{um-&-up}
u_{p}\, = \, \frac{1}{2} \, \left[ - \, \frac{\beta_4}{\beta _6} \, + \, 
\sqrt{\left(\frac{\beta_4}{\beta _6}\right)^2 \, - \, 4 \, \frac{\beta_2}{\beta _6}} \right] 
\quad \& \quad 
u_{m}\, = \, - \, \frac{1}{2} \, \left[ \frac{\beta_4}{\beta _6} \, +\, 
\sqrt{\left(\frac{\beta_4}{\beta _6}\right)^2 \, - \, 4 \, \frac{\beta_2}{\beta _6}} \right] \, . 
\end{equation}
In the infinite size limit, where $\beta_6<0$, $\beta_4>0$ and $\beta_2 = 0$, the bending point of the string is at 
$u_{m}=0$.


\subsection{Boundary conditions}

In this subsection we will impose boundary conditions to the equations of motion of the previous subsection,
in analogy to the infinite size case of \cite{Georgiou:2017pvi}. 

Solving \eqref{constraintZ0} with respect to $A_T$ we obtain the following expression
\begin{equation} \label{ATgeneral}
A_T \, = \, \frac{1}{d\, Z_0^2} \, \Big( 2 \, A_{\phi} \, c  \, \mu \, - \, \alpha \,  c^2 \, 
+ \, d^2 \, \kappa \, Z_0^2 \Big) \, . 
\end{equation}
Substituting \eqref{ATgeneral} in \eqref{constraintV} and imposing that $V_y'$ vanishes at the bending point 
$u = u_{m}$ we obtain the following two solutions for the constant $A_{\phi}$
\begin{eqnarray}
\label{AGM}
& &
A_\phi\,=\,\frac{\alpha}{2 \, c\, \kappa \, \mu^2}\,
\Big[ c^2  \, \kappa \,  \mu \, + \, c \,  d \, \omega \,  Z_0^2 \, + \, d^2 \, \kappa \,  \mu \Big]
\hspace{1.35cm} \textrm{single spin giant magnon}
\\
\label{ASS}
\rule{0pt}{.9cm}
& &
A_\phi\,=\,\frac{c\, \alpha}{ 2 \, \mu}
\hspace{6.8cm} \textrm{single spin single spike} \, . 
\end{eqnarray}

Substituting the expression for $A_T$ and $A_\phi$ (from \eqref{ATgeneral} and \eqref{AGM} respectively) 
in \eqref{um-&-up}, we express $u_m$ (of the giant magnon) in terms of the auxiliary quantity $W_{GM}$, as follows
\begin{equation} \label{um+WGM}
u_m \, = \, 1 -W_{GM} \quad {\rm with} \quad W_{GM}\, = \, \frac{\alpha \, Z_0^2}{\kappa \, \mu^2} \, .
\end{equation}
The expression for $u_p$ (of the giant magnon) becomes
\begin{equation}  \label{up+DGM}
u_p\, = \, 1-v^2 \, W_{GM} \Bigg[ 1+ \frac{1-v^2}{v} \, \frac{\Delta_{GM}}{W_{GM} +2 \, v\, \Delta_{GM}} \Bigg]^2
\quad {\rm with} \quad
\Delta_{GM} \, = \,  \frac{ \alpha}{\mu \, \omega}
\quad \& \quad 
v \, = \, \frac{d}{c} \, . 
\end{equation}
The expressions for $Z_0$ and $\kappa$ in terms of the auxiliary quantities $W_{GM}$ and $\Delta_{GM}$ are
\begin{equation}
Z_0 \, =  \, \sqrt{\frac{\kappa}{\alpha}} \, \mu \, \sqrt{W_{GM}}
\quad \& \quad 
\kappa\, = \, \omega \, \Bigg[W_{GM} +2 \, v\, \Delta_{GM} + \left(1+v^2\right) \, 
\frac{\Delta_{GM}^2}{W_{GM}}\Bigg]^{1/2} \, .
\end{equation}
Notice, that the undeformed limit of the expressions above is for $\Delta_{GM} \rightarrow 0$ 
(or equivalently $\alpha \rightarrow 0$) and the infinite size limit is for $W_{GM} \rightarrow 1$.

A similar analysis can be performed also for the single spin single spike case. Now the starting point for the value 
of the constant $A_\phi$ will be \eqref{ASS}. The expression for 
$u_m$ (of the single spike) in terms of the auxiliary quantity $W_{SS}$ becomes
\begin{equation} \label{um+WSS}
u_m \, = \, 1 -W_{SS} \quad {\rm with} \quad 
W_{SS} \, = \, \frac{\alpha}{\mu \, v } \, \left(\omega + \frac{2 \, \kappa \, \mu}{v \, Z_0^2}\right)^{-1}\, .
\end{equation}
The expression for $u_p$ (of the single spike) becomes
\begin{equation}
u_p\, = \, 1-v^2 \, W_{SS} \quad {\rm with} \quad v \, = \, \frac{c}{d}
\end{equation}
and for later convenience we introduce the following auxiliary quantity $\Delta_{SS}$
\begin{equation} \label{DSS}
\Delta_{SS} \, = \,  \frac{\mu \, \omega}{ \alpha} \, . 
\end{equation}
The expressions for $Z_0$ and $\kappa$ in terms of the auxiliary quantities $W_{SS}$ and $\Delta_{SS}$ are
\begin{equation}
Z_0 =  \,  \frac{\sqrt{2 \, \kappa \, \alpha \, W_{SS}} \,  \Delta_{SS}}{\omega \, 
\sqrt{1 - v \, W_{SS} \, \Delta}} 
\quad \& \quad 
\kappa \, = \, \frac{\omega}{\sqrt{W_{SS}} \, \Delta_{SS}} \, .
\end{equation}
Notice, that the undeformed limit of the expressions above is for $\Delta_{SS} \rightarrow \frac{1}{v \, W_{SS}}$ 
(or equivalently $Z_0 \rightarrow \infty$) and the infinite size limit is for $W_{SS} \rightarrow 1$.


\section{Single Spin Giant Magnon}
\label{section-GM}

In this section we compute the conserved charges for the single spin giant magnon solution and construct the dispersion 
relation. The detailed analysis for the calculation of the dispersion relation in the infinite size limit appears in 
\cite{Georgiou:2017pvi, Georgiou:2018zkt}. In this section we focus on the finite size corrections of the 
aforementioned relation. 

The four conserved charges, that originate from the partial derivatives of the Polyakov action and the subsequent integration, 
for the giant magnon solution  become\footnote{In order for the notation not to clutter, in this section $W$ is $W_{GM}$ from 
 \eqref{um+WGM} and $\Delta$ is $\Delta_{GM}$ from \eqref{up+DGM}.}
\begin{eqnarray}
\label{EnergyGM}
 \frac{E}{2 \, T}&=& \frac{1-v^2}{\sqrt{u_p}} \, \frac{W}{W +2 \, v\, \Delta} \, 
 \Bigg[W +2 \, v\, \Delta + \left(1+v^2\right) \, \frac{\Delta^2}{W}\Bigg]^{1/2}\, 
 \, \mathbf{K}(1-\epsilon)
\\
\label{MGM}
\rule{0pt}{.8cm}
\frac{\mu \, M}{2 \, T} &=&  \frac{1-v^2}{\sqrt{ u_p}} \,  \, \frac{\Delta}{W +2 \, v\, \Delta} \, 
\, \, \mathbf{K}(1-\epsilon)
\\
\label{JpsiGM}
\rule{0pt}{.8cm}
\frac{J}{2 \, T} &=&  \frac{1}{\sqrt{ u_p}} \, \Bigg[1- \frac{v^2 \, W}{W +2 \, v\, \Delta}\left(W + \frac{1+v^2}{v}\, \Delta\right)\Bigg] \, \, \mathbf{K}(1-\epsilon) - 
\sqrt{ u_p} \, \, \, \mathbf{E}(1-\epsilon)
\\
\label{DeltaGM}
\rule{0pt}{.8cm}
\frac{\Delta \varphi}{2}
&=&  \frac{v \,W + \left(1+v^2\right) \, \Delta}{\left(W +2 \, v\, \Delta\right) \, \sqrt{u_p}}\,
\Bigg[ \frac{W}{1-u_p} \, \mathbf{\Pi}\left(-\frac{u_p}{1-u_p}\, \left(1-\epsilon\right),1-\epsilon\right)  - 
\mathbf{K}(1-\epsilon) \Bigg]
\end{eqnarray}
where $\epsilon$ is the ratio between $u_{m}$ and $u_{p}$ of the giant magnon solution 
\begin{equation} \label{def-epsilon}
\epsilon \, = \, \frac{u_{m}}{u_{p}}
\end{equation}
and $ \mathbf{K}(1-\epsilon)$, $ \mathbf{E}(1-\epsilon)$ and 
$\mathbf{\Pi}\left(-\frac{u_p}{1-u_p}\, \left(1-\epsilon\right),1-\epsilon\right)$ 
are the complete elliptic integrals of the first, the second and the third kind. In appendix \ref{appendix-2}
we have gathered all the expressions that define those integrals. The finite size corrections we will calculate are in 
powers of the ratio $\epsilon$ when this ratio is small. For this reason we expand the parameters $v$ and $W$ in powers
of $\epsilon$ as follows
\begin{equation} \label{v-W-expansion}
v \, = \, v_0 + \left(v_1 + v_2 \, \log \epsilon \right) \epsilon 
\quad \& \quad
W \, = \, W_0 + W_1 \, \epsilon
\end{equation}
and the presence of the logarithmic term is for the  compensation of the logarithmic terms that come 
from the expansion of the elliptic integrals. Using the definition for $\epsilon$ from \eqref{def-epsilon} 
and the condition $\Delta \varphi = p$, it is possible to determine all the coefficients of the above expansion. 
The coefficients of the $W$ expansion are simple expressions and do not depend on the deformation parameter $\Delta$
\begin{equation}
W_0\, =\, 1 \quad  \& \quad W_1 \, = \, -\sin^2\frac{p}{2} \, . 
\end{equation}
On the contrary, the coefficients of the $v$ expansion depend on the deformation parameter and they have analytic 
but non illuminating expressions. We list them in appendix \ref{appendix-2} and more specifically in equations 
\eqref{v0-GM-full}, \eqref{v2-GM-full} and \eqref{v1-GM-full}. Here we present the expansion of the coefficients around the 
undeformed values, namely for small values of $\Delta$. For the coefficient $v_0$ the expansion becomes
\begin{equation}
\frac{v_0}{ \cos \frac{p}{2}} \, = \, 1 - \frac{\sin^2 \frac{p}{2}}{\cos \frac{p}{2}} \, \Delta 
+ {\cal O} (\Delta^3)
\end{equation}
for the coefficient $v_1$ it is 
\begin{equation}
\frac{v_1}{\frac{1}{4} \, \sin^2 \, \frac{p}{2}\cos \, \frac{p}{2} \, (1 - 4 \, \ln 2)} \, = \, 1 +  
\frac{2 - 6 \, \sin^2 \, \frac{p}{2} - 8 \, \cos^2 \, \frac{p}{2} \, \ln 2}{\cos \, \frac{p}{2}  \left(1 - 4\, \ln 2\right)} \, \,  \Delta 
+  {\cal O} (\Delta^3)
\end{equation}
and for the coefficient $v_2$ it is 
\begin{equation}
\frac{v_2}{\frac{1}{4} \, \sin^2 \, \frac{p}{2}\cos \, \frac{p}{2}} \, = \, 1 + 2 \, \cos \frac{p}{2} \, \Delta 
+  {\cal O} (\Delta^3) \, . 
\end{equation}
Notice here, that the current calculation of finite size corrections in the dispersion relation of the giant magnon only makes 
sense in the limit of small $\Delta$. Increasing the value of the deformation parameter, increases the ratio between $u_m$
and $u_p$ and as a result cancels the expansion in small $\epsilon$. The expansion is justified only for small values of $\Delta$.

From the zeroth order term in the expansion of $J$, we obtain the expression of $\epsilon$ for $J\gg T$
\begin{equation}
\epsilon \, = \, 16 \, {\rm exp} \left[{\left[- \frac{J}{T\, \sin \frac{p}{2}}-2\right] \, \left[ 1 - \Delta \, \cos \frac{p}{2} \right]}\right] 
\end{equation}
where we have expanded the exponent in powers of $\Delta$.

Using all those ingredients it is possible to expand the dispersion relation of the giant magnon. 
The zeroth order term provides the infinite size result while the first order correction depends on 
the value of $\Delta$. For small values of $\Delta$ the dispersion relation is
\begin{eqnarray} 
\frac{\sqrt{\big.E^2  \, - \, \mu^2 \, M^2} \, - \, J }{2 \, T\,  \sin  \frac{p}{2}} -1 &=& 
-4\, \sin^2 \frac{p}{2} \, e^{- \frac{J}{T\, \sin \frac{p}{2}}-2}
\\
&+&
2\, \Delta^2
\Bigg[ \frac{J^2}{T^2}\, \cos^2 \frac{p}{2} + 2 \, \frac{J}{T} \, \cos p \, \sin \frac{p}{2} - 4\, \sin^4 \frac{p}{2}\Bigg] 
e^{- \frac{J}{T\, \sin \frac{p}{2}}-2} \, .
\nonumber
\end{eqnarray}
For $\Delta =0$ we obtain the undeformed result of \cite{Arutyunov:2006gs} (see also \cite{Ahn:2008sk}) and in the 
infinite size limit the RHS of the finite size dispersion relation vanishes.


\section{Single Spin Single Spike}
\label{section-SS}

In this section, we compute the conserved charges for the single spin single spike solution and construct the dispersion 
relation. The detailed analysis for the calculation of the dispersion relation in the infinite size limit appears in 
\cite{Georgiou:2017pvi, Georgiou:2018zkt}. In this section we focus on the finite size corrections of the 
aforementioned relation.

The four conserved charges for the single spike solution become\footnote{In this section $W$ is $W_{SS}$ from 
 \eqref{um+WSS} and $\Delta$ is $\Delta_{SS}$ from \eqref{DSS}.}
\begin{eqnarray}
\label{EnergySS}
 \frac{E}{2 \, T}&=& \, \sqrt{\frac{1}{v^2}-1} \,\, \sqrt{1 - \epsilon}
 \, \mathbf{K}(1-\epsilon)
\\
\label{MSS}
\rule{0pt}{.8cm}
\frac{\mu \, M}{2 \, T} &=&  \, \frac{1}{2} \, \left(1- v \, W \, \Delta \right) \, \sqrt{\frac{1}{v^2}-1} \,\, \sqrt{1 - v^2 \, \epsilon}
\, \, \mathbf{K}(1-\epsilon)
\\
\label{JpsiSS}
\rule{0pt}{.8cm}
\frac{J}{2 \, T} &=&  \sqrt{\frac{1 - v^2}{1 - v^2 \, \epsilon}}\, \, \Bigg[ \mathbf{E}(1-\epsilon) - \epsilon \, \mathbf{K}(1-\epsilon) \Bigg]
\\
\label{DeltaSS}
\rule{0pt}{.8cm}
\frac{\Delta \varphi}{2}
&=& \sqrt{ \frac{1 - v^2 \, \epsilon}{1 - v^2} }\,
\Bigg[ \left[v + \frac{1-v^2}{2\, v} \, \left(1- v \, W \, \Delta \right)\right]\, \mathbf{K}(1-\epsilon)- 
\,\frac{1}{v} \, \mathbf{\Pi}\left( 1 - \frac{1}{v^2},1-\epsilon\right) \Bigg]
\end{eqnarray}
where the constant $W$ can be expressed as a function of $\epsilon$ as follows
\begin{equation}
W \, = \, \frac{1- \epsilon}{1 - v^2 \, \epsilon}
\end{equation}
and $\epsilon$ is again the ratio between $u_m$ and $u_p$ (of the single spike).
The next step is to expand $v$ in powers of $\epsilon$ (see equation \eqref{v-W-expansion}), substitute 
it in the expression for $J/T$ from \eqref{JpsiSS} and impose that $J/T$ remains finite.
In that way, we determine the value of the coefficients $v_0$, $v_1$ and $v_2$ in terms of the finite 
quantity $J/T$
\begin{equation}
v_0^2 \, = \, 1 - \frac{1}{4} \, \frac{J^2}{T^2}\, , \quad 
v_1 \, = \, - \, \frac{1 - v_0^2}{4\, v_0} \, \Big[1 + 4 \, \ln 2- 2\, v_0^2\Big] 
\quad \& \quad
v_2 \, = \,  \frac{1 - v_0^2}{4\, v_0} \, .
\end{equation}
From the $\epsilon$-expansion of $\Delta \varphi$, we obtain the expression for $\epsilon$ as a 
function of $\Delta \varphi$ and $J/T$
\begin{equation}
\epsilon \, = \, 16 \, {\rm exp} \left[ - \frac{4\, \sqrt{\big.4-{\cal J}^2}}{{\cal J}} \, \frac{\Delta \varphi + 
\arcsin \left(\frac{\cal J}{2} \, \sqrt{\big.4-{\cal J}^2}\right)}{2 + \Delta \,\sqrt{\big.4-{\cal J}^2}} \right]
\quad {\rm with} \quad
{\cal J} = \frac{J}{T} \, . 
\end{equation}
Using all those ingredients it is possible to expand the dispersion relation of the single spike.
The zeroth order term provides the infinite size result while the first order correction depends on 
the value of $\Delta$. Here contrary to the giant magnon result the expressions are simple and we do not expand 
in powers of the deformation. The finite size dispersion relation is
\begin{equation}
\frac{1}{2 \, T} \, \Big[ E - \mu \, M - \Delta \varphi \Big] - \frac{p}{2} = 4 \, \sin^2 \frac{p}{2} \, \tan \frac{p}{2}
\,  {\rm exp} \left[-\, \frac{\Delta \varphi + p}{\tan \frac{p}{2}} \, \frac{2}{1 +\Delta \, \cos \frac{p}{2} }\right]
\end{equation}
where we have used the identification that was introduced in \cite{Ishizeki:2007we}
\begin{equation}
\arcsin \left(\frac{J}{2 \, T}\right) \, = \, \frac{p}{2} \, . 
\end{equation}
The undeformed result of \cite{Ahn:2008sk} is realized for $\Delta = \cos^{-1}\frac{p}{2}$ (this is the same value of $\Delta$ 
for which $Z_0 \rightarrow \infty$) and in the 
infinite size limit the RHS of the finite size dispersion relation vanishes.

\section{Conclusions}
\label{conclusions}

In this paper we have elaborated on the dispersion relations of semi-classical string solutions that live in the 
Schr\"odinger spacetime, that is conjectured to be the gravity dual of the null dipole CFT. 
We have calculated the finite size corrections to the infinite size dispersion relation of the single spin giant magnon 
and of the single spin single spike solutions. The leading order corrections have the usual exponential form, characteristic 
for the finite size corrections also in the original $AdS_5 \times S^5$ background, dressed with expressions that depend 
on the deformation parameter. In the limit of zero deformation, the solutions and their 
finite size dispersion relations become those of the single spin giant magnon and single spin single spike of the 
undeformed $AdS_5 \times S^5$ background. Furthermore, in the infinite size limit the correction term vanishes and we obtain 
the known infinite size dispersion relations. Even if the solutions presented in this paper are single spin they live in an $S^3$ 
subspace of the five-sphere. In order to have a dispersion relation that depends on the deformation, we are forced to let the string 
move inside an $S^3$, instead of the $S^2$ that the string would move in the undeformed counterpart.

A number of important directions remains to be addressed. In the current paper we have calculated the finite size corrections 
for the single spin giant magnon and single spike. It would be interesting to generalize the computation for the 
dyonic giant magnon and single spike. The string solutions and the corresponding infinite size dispersion relations 
have been studied in \cite{Georgiou:2017pvi}. In a very recent interesting approach, the classical exponential corrections to 
the dispersion relations of the GKP string, of the giant magnon and of the single spike have been expressed 
in terms of Lambert W-function \cite{Georgiou:2010zt,Floratos:2013cia, Floratos:2014gqa}.  In this way the leading, sub-leading 
and next-to-sub-leading series of the classical exponential corrections to the dispersion relations of the aforementioned
string configurations have been calculated. It would be very interesting to follow this path in the 
Schr\"odinger geometry in order to expand the computation of the current paper beyond the leading order.


\section*{Acknowledgments}

We would like to thank George Georgiou for reading very carefully
the preliminary draft and sending various useful comments.
We are grateful to Georgios Linardopoulos for collaboration in the early stages of this work.
The work of this project has received funding from the Hellenic Foundation
for Research and Innovation (HFRI) and the General Secretariat for Research and Technology (GSRT), under grant agreement No 15425.


\appendix


\section{Useful expressions from the analysis of the classical string solutions}
\label{appendix-1}

In this appendix we gather all the useful (but lengthy and not particularly illuminating) expressions 
from the detailed analysis of the classical string solutions, that is presented in section \ref{ClassicalSolutions}.

The derivative of the function $V_y(y)$ with respect to $y$ in terms of $u(y)$ is given by
\begin{equation} \label{constraintV}
V_y'(y) \, = \, \frac{1}{c^2 \, - \, d^2} 
\Bigg[ \left( d \, \kappa  - A_T\right) Z_0^2 - 
\alpha \, d  +  c \, \mu \, \omega \, \left(1 + \frac{2\, d\, \kappa \, \mu}{c \, \omega \, Z_0^2} \right) \,
\left(1 \, - \, u(y) \right) \Bigg]
\end{equation}
while the derivative of the function $\Psi_y(y)$ with respect to $y$ is given by
\begin{equation} \label{constraintPsi}
\Psi_y'(y) \, = \, \frac{1}{c^2 \, - \, d^2} 
\Bigg[ \frac{2 \, A_{\phi}}{1\,- \, u(y)}  -  d \, \omega  - \frac{2 \, c \, \kappa \, \mu }{Z_0^2}\Bigg] \, . 
\end{equation}
The constants $\beta_2$, $\beta_4$ \& $\beta_6$ from the equation of motion for the function $u(y)$ 
in \eqref{equation-u} are given by the following expressions
\begin{eqnarray}
\label{beta 4}
\hspace{-.7cm}
\beta_4 \,& = & \, \frac{1}{c\, d\, Z_0^2 \,(c^2-d^2)^2}  \Bigg[
- 2 \, A_{\phi} \, \left(c\, Z_0^2 \, \omega + d \, \kappa \, \mu \right) \,\left(c^2 \, + \, d^2 \right) - 
4 \, A_{\phi}  \, c^2 \, d \, \kappa \, \mu   
\\
\hspace{-.7cm} 
&+& c \, \kappa  \left(c^2 - d^2 \right) \left[Z_0^2 \left(A_T - d \, \kappa \right) + \alpha  \, d \right]
+ \frac{2 \, c \, d}{Z_0^2} \, \left(c\, Z_0^2 \, \omega + 2 \, d \, \kappa \, \mu \right)^2 \Bigg] 
\nonumber \\
\label{beta 6}
\hspace{-.7cm}
\beta_6 \,& = & \, - \,  \frac{c^2 \, \omega^2 }{\left(c^2\,- \, d^2 \right)^2} \, 
\left(1 + \frac{2\, d\, \kappa \, \mu}{c \, \omega \, Z_0^2} \right)
\, < \, 0 
\quad \& \quad 
\beta_2 + \beta_4 + \beta_6 = - \frac{4 \, A^2_{\phi}}{(c^2-d^2)^2} \, . 
\end{eqnarray}


\section{Elliptic integrals and $\epsilon$ expansion coefficients of the parameter $v$}
\label{appendix-2}

In this appendix we gather all the useful expressions for the definitions of the elliptic integrals (of the first, second and third type)
that we used to express the conserved quantities, both for the giant magnon and single spike solutions.  
These are
\begin{eqnarray}
&& \int_{u_m}^{u_p} \, \frac{du}{ \sqrt{u \, \left( u_{p} \, - \, u \right) \, \left(u \, - \, u_{m} \right)}} \, = \, 
\frac{2}{\sqrt{u_{p}}} \,\mathbf{K}(1-\epsilon) 
\\[7pt]
&& \int_{u_m}^{u_p} \,\frac{\sqrt{u} \, \,du}{ \sqrt{\left( u_{p} \, - \, u \right) \, \left(u \, - \, u_{m} \right)}} \, = \, 
2 \, \sqrt{u_{p}} \,\mathbf{E}(1-\epsilon) 
\\[7pt]
&&  \int_{u_m}^{u_p} \, \frac{du}{\left(1 - u\right) \sqrt{u \, \left( u_{p} \, - \, u \right) \, \left(u \, - \, u_{m} \right)}} \, = \, 
\frac{2}{\left(1 - u_{p} \right)\sqrt{u_{p}}} \,\mathbf{\Pi}\left(\frac{u_{m} -u_{p}}{1- u_{p}} |1-\epsilon\right)  \, . 
\end{eqnarray}
The expressions for the $\epsilon$ expansion coefficients of the parameter $v$ in the single spin giant magnon solution of 
section \ref{section-GM} are 
\begin{equation} \label{v0-GM-full}
v_0 \, = \, \frac{1}{2 \, \Delta} \left[\sqrt{1+4 \, \Delta \Bigg[\Delta \, \cos p + \cos \frac{p}{2} \sqrt{1- 4\, \Delta^2 \, \sin^2 \frac{p}{2}}\Bigg]}-1\right]
\end{equation}
\begin{equation} \label{v2-GM-full}
v_2 \, = \, \frac{\left(1-v_0^2\right) \left[v_0 + \left(1+v_0^2\right) \, \Delta \right]}{4 \, \left(1 +2 \, v_0 \, \Delta \right)} \, 
\left[1 +\frac{\left(1-v_0^2\right) \Delta^2}{1+2 \, \Delta \, \Big[v_0 - \Delta \left(1-v_0^2\right)\Big]}\right]
\end{equation}

\begin{equation} \label{v1-GM-full}
\frac{v_1}{v_2} \, = \, \frac{v_0 \, \left(1-4 \, \ln 2\right)\, \left[1 +2 \, \Delta^2 \, \left(1+v_0^2\right) \right] - 
\Delta \, \left(3+4\, \ln 2\right) + v_0^2 \, \Delta \, \left(7 -12 \, \ln 2\right)}{ 
\left(1 +2 \, v_0 \, \Delta \right) \left[v_0 + \left(1+v_0^2\right) \, \Delta \right]} \, . 
\end{equation}


\end{document}